\documentclass[12pt]{amsart}

\usepackage{etex}
\usepackage{amscd}
\usepackage{amsmath}
\usepackage{amsfonts}
\usepackage{amssymb}
\usepackage{graphics}
\usepackage{epsfig}
\usepackage{pictex} 

\usepackage{booktabs}       
\usepackage{lipsum}
\usepackage{graphicx}
\usepackage{float, tabularx}
 2

\numberwithin{equation}{section}
\numberwithin{theorem}{section}

\title{Development of Choice Model for Brand Evaluation}

\author{Marina Kholod}
\address{Plekhanov Russian University of Economics
\newline \indent 115054, Moscow, Russia}
\email{kholod.mv@rea.ru}

\author{Nikita Mokrenko}
\address{Plekhanov Russian University of Economics
\newline \indent 115054, Moscow, Russia}
\email{mokrenko.nv@rea.ru}

\thanks{This research was funded by the Ministry of Science and Higher Education of the Russian Federation, grant number FSSW-2023-0004.}

\begin{document}

\begin{abstract}
Consumer choice modeling takes center stage as we delve into understanding how personal preferences of decision makers (customers) for products influence demand at the level of the individual. The contemporary choice theory is built upon the characteristics of the decision maker, alternatives available for the choice of the decision maker, the attributes of the available alternatives and decision rules that the decision maker uses to make a choice. The choice set in our research is represented by six~major brands (products) of laundry detergents in the Japanese market. We use the panel data of the purchases of 98 households to which we apply the hierarchical probit model, facilitated by a Markov Chain Monte Carlo simulation (MCMC) in order to evaluate the brand values of six~brands. The applied model also allows us to evaluate the tangible and intangible brand values. These evaluated metrics help us to assess the brands based on their tangible and intangible characteristics. Moreover, consumer choice modeling also provides a framework for assessing the environmental performance of laundry detergent brands as the model uses the information on components (physical attributes) of laundry detergents.  Through a comprehensive evaluation of product performance, including brand tangible estimation, we shed light on the sustainability attributes of laundry detergents, offering a roadmap for consumers and manufacturers alike to make more informed, environmentally responsible choices of laundry detergents based on their physical attributes. Knowing the estimates of the attributes for the laundry detergent products, manufacturers can modify their physical attributes, e.g., decrease the amount of the detergent needed for one wash while increasing the total weight of the laundry powder in the package. In this way, more ecology- and consumer-friendly decisions can be made by manufacturers of laundry detergents.

\end{abstract}

\maketitle

\section{Introduction
\label{sec:introduction}}

As global awareness of our collective environmental impact grows, sustainable practices are becoming increasingly important, while it may be easy to overlook laundry detergents, these everyday cleaning agents can make a significant difference in the eco-friendliness of our daily routines. By understanding the environmental implications of traditional laundry detergents and making conscious choices in selecting and using alternatives, we can significantly reduce our ecological footprint.

Traditionally laundry detergents impact the environment through release of chemical components, non-biodegradable ingredients, and plastic packaging.  As for chemical components, conventional laundry detergents often contain harmful chemicals, such as phosphates, fragrances, and optical brighteners. These toxic ingredients can leach into water supplies, causing harm to aquatic life and contributing to environmental pollution. As for non-biodegradable ingredients,  many traditional laundry detergents contain non-biodegradable substances, such as synthetic surfactants. When discharged into the environment, these substances accumulate over time, posing potential threats to ecosystems and public health. As for plastic packaging, the plastic bottles commonly used to package laundry detergents are composed of non-renewable resources, like petroleum, and contribute to plastic pollution. Improper disposal leads to the contamination of waterways, marine life, and the wider environment. {Adopting sustainable laundry practices, such as using energy-efficient machines and eco-friendly detergents, can significantly reduce the environmental impact of clothing care. Recent research~\cite{mii,laitala} underscores the importance of these practices, highlighting their potential to mitigate water and energy consumption, thereby promoting a more environmentally conscious approach to laundry routines. Tomsic, Ofentavšek and Fink~\cite{tomsic} investigate the use of different types of laundry detergent at different washing temperatures and suggest using powders with a certain composition. Nawaz and Sengupta~\cite{nawaz} analyze the role of each detergent component in affecting the environment. Järvi and Paloviita~\cite{jarvi} examine how households read, comprehend, and adhere to the directions on detergent packages and dosage guidelines.}

In our work, we take a consumer perspective, for which the product name can be defined as the collection of the concepts that a consumer learns to associate with a certain product (represented by the product). It is very obvious in this context 
that product value has many implications when there is an extension of one product to many under different product names.
The measurement of product equity can be carried out through the utility and preferences that the consumer attaches to the product name within the framework of choice model.

We divided all studies about the value of the product to the consumer into two large groups. The first group is represented by research about contribution of producting to physical products, such as Tauber~\cite{tauber}, Aaker and Keller~\cite{aaker}, and Schlossberg~\cite{schlossberg}. These researchers consider the product name as a collection of concepts. They also study the associations of consumers in response to product name and finally talk about the product extension across several products. They extend the idea of product formation of Levitt and see the  product name as a multi-faceted symbol that consumers come to associate with specific concepts through their interactions and experiences with that product. According to Aaker and Keller~\cite{aaker}, when judging the launch of a new product extension, consumers are likely to be more accepting if they perceive a degree of congruence in terms of product compatibility or shared attributes. That is to say, if a company, already known for a specific product type or set of values, offers a new product that aligns closely with that existing image, consumers are apt to respond more positively. When a corporation grows its product line, it leverages these familiar brand names across new product categories within their portfolio, effectively maximizing their most significant assets---consumer recognition, positive sentiment, and the impressions linked to the brand name.

The second group is represented by researchers such as Louviere and Johnson~\cite{louviere:pr_equity}, Sharkey~\cite{sharkey} who measure the value or usefulness that consumers assign to product names directly. This is achieved through conjoint analysis where consumers evaluate combinations of product attributes and product names. The exposed preferences are then disassembled into two parts: the utility linked to the product attributes and the value associated with the product names. Tybout and Hauser~\cite{tybout} use the model of consumer decision making developed by Hauser and Urban~\cite{hauser} and apply it to the marketing audit for mature product with the purpose of identifying further actions that can increase profitability. Shocker and Srinivasan~\cite{schoker} review main models existing at that time with the purpose of increasing implementability of marketing concepts, permitting consumer inputs to be used at the earliest stages in the development and selection of marketing strategies. Kamakura and Russell~\cite{kamakura} adopt the theoretical model of consumer decision making discussed by Tybout and Hauser. Andrews, Luo, Fang, and Ghose~\cite{andrews} explored how online channels can be utilized to measure and influence brand equity. These studies often employ conjoint analysis, wherein consumers rate combinations of product features and brand names, with the resulting preferences being broken down into utility linked to product features and value assigned to brand names.


\section{Materials and Methods}
\label{sec:mndm}

The initial stage starts with the evaluation of a product's physical attributes and the psychological and social aspects of the consumer. Tybout and Hauser suggest that Brunswik's model can be utilized to understand the connection between these attributes and the resultant subjective assessments or perceptions~\cite{brunswik1}.  According to this model, physical features r of the product, such as (in our case) concentration of surface-active agent (S.A.A.), presence of bleach in the laundry detergent, as well as the type of package, the amount of detergent needed for 30 L of water, and net weight of the detergent in the package, form the fundamentals for consumers’ perception (${Y}_{h}$) of the detergent washing power of the detergent product ${j}$. A physical feature does not necessarily lead to a unique perception but contributes to the different perceptions in different ways. Determining the effect of non-physical aspects of a product on consumer perceptions can be challenging due to the sheer volume of such cues and their subjectivity at the individual level.

A product  ${j}$ has ${R}$ physical features ${D}_{jr}\:(r = 1,2,\ldots,R)$. For a consumer ${h}$, their perception is formulated as a function of product ${j}$ physical features and consumer ${h}$ psychosocial cues. Hence, the perception ${Y}_{hjq}$ for the attribute ${q\:(q = 1,2,\ldots,Q)}$ by consumer ${h}$ about product ${j}$ is

\begin{equation}
Y_{hjq}= \sum_{r=1}^{R} w_{hrq} D_{jr} + v_{hjq},
\end{equation}

where $Y_{hjq}$ is the attribute perception for product $j$ on an attribute $q (q = 1,2,\ldots,Q)$ for a consumer $h$, $D_{jr}$ is a product $j$’ s actual physical features $R$, $w_{hrq}$---coefficients that represent the $R$ physical characteristics of product $j$ into its perceived attributes in a $Q$-dimensional attribute space---and $v_{hjq}$ is an error due to the perceptional distortions which arise in response to the psychosocial cues. 

So, consumers form perceptions based on a combination of physical attributes\linebreak ($\sum_{r=1}^{R} w_{hrq} D_{jr}$) and a distortion of these attributes ($v_{hjq}$). For example, consumers might form their perception of “washing power” attribute of a laundry detergent based upon the information that the consumer has about the physical features of a product (e.g., the amount of S.A.A. and bleach needed to wash dirty clothes) and upon advertisement of special cleaning substances that the product has. 

In the second step, after collecting (which was performed in the initial step) all the attribute perceptions $Y_{hjq}\:q (q = 1,2,\ldots,Q)$, the consumer is going to order “by preference” these attributes using a weight factor $\theta_{q}$. So, we can (by addition) obtain the preferences of product $j$ by consumer $h$ as such:

\begin{equation}
P_{hj}=\sum_{q=1}^{Q} \theta_{hq}Y_{hjq}+\varphi_{hj},
\end{equation}

where $P_{hj}$ are the preferences for product $j$ by a consumer $h$, $\theta_{hq}$ is the relative importance assigned to each perceived attribute $q$ by a consumer $h$, and $\varphi_{hj}$ is a factor in the preferences which is not contained in the attribute perceptions $Y_{hjq}$ of product $j$.

These importance weights $\theta_{hq}$ reflect how consumer $h$ translates their perceptions of the available products into preferences.

At this stage, the “engineering parameters”, which relate the physical characteristics of the product to the consumer’s evaluation of the product, are denoted as 

\begin{equation}
\delta_{hr}=\sum_{q=1}^{Q} w_{hrq}\theta_{hq}
\end{equation}

The coefficients $\delta_{hr}$ are not attribute importance weights.

The intangible part of the product’s value, which arises from different product associations and perceptual distortions, is denoted as

\begin{equation}
\varphi^{*}_{hj}=\varphi_{hj}+\sum_{q=1}^{Q} (\theta_{hq}v_{hjq}),
\end{equation}

where $\varphi_{hj}$  is a factor in the preferences which is not contained in the attribute perceptions $Y_{hjq}$ of product j and $\sum_{q=1}^{Q} (\theta_{hq}v_{hjq})$ is a factor in the preferences, which is based on importance weights of perceived attribute, such as “washing power”, “bleaching power”, and “amount of foam”, and perceptional distortions that emerge as a reaction to psychosocial signals, for example, to the advertising about those attributes. 

As a result, we obtain a product $j$’s value for a consumer $h$ decomposed into a tangible component ($\sum_{r=1}^{R} \delta_{hr}D_{jr}$), which is directly linked to the physical characteristics of the product, and an intangible part ($\varphi^{*}_{hj}$), which emerges from misconceptions and other associations related to the product.

\section{Results}

In the first stage of our data analysis, we estimated the parameters of the multinomial model.

\begin{table}[H] 
\caption{Detergent Attribute Data\label{tab3}}
\newcolumntype{C}{>{\centering\arraybackslash}X}
\begin{tabularx}{\textwidth}{CCCCCC}
\toprule
\textbf{}	& \textbf{Posterior Mean}	& \textbf{S.D.} & \textbf{HPD} & \textbf{(+)} & \textbf{(-)}\\
\midrule
\multicolumn{6}{c}{Market Response Parameter} \\
Display    & 1.523   & 0.618  & 98  & 96 & 2  \\
Price      & -4.331  & 0.639  & 98  & 0  & 98 \\
Product 1  & 2.514   & 1.676  & 96  & 89 & 7  \\
Product 2  & 2.188   & 1.846  & 97  & 87 & 10 \\
Product 3  & 1.313   & 1.674  & 98  & 79 & 19 \\
Product 4  & 1.173   & 2.258  & 97  & 67 & 30 \\
Product 5  & 1.529   & 1.939  & 97  & 78 & 19 \\
Product 6  & 1.358   & 1.659  & 95  & 78 & 17 \\
\bottomrule
\end{tabularx}
\end{table}

Table~\ref{tab3} shows that the posterior mean of consumer’s parameter estimates, the standard deviation and the number of consumers with statistically significant estimate tested by 95\% HPD regions (the last two columns denote respectively the number of consumers with positive and negative signs of their parameters estimates). We can stipulate that all the parameters estimates are statistically significant for almost all the consumer’s sample. All of the 98 consumers have statistically significant price estimates and react negatively, as expected, to the price. The same results are obtained regarding the Display parameters with 2 consumers having a negative response to display but those estimates are not significant.

As for the intercept parameters, a similar conclusion can be made regarding the statistical significance of their estimates over the whole sample of consumers. However, negative estimated values are recorded for a number of consumers, for instance 30 units, over 25\% of the sample size, have a negative value of the intercept estimate for Product 4. This result might be linked, to a certain extent, to the difference in the behavior of consumers regarding their choice formation.

As for the Products, the largest estimate of intercepts is 2.514, for Product 1. The smallest is 1.173 for Product 4, which is 2.14 times of Product 1.

Table~\ref{tab4} shows that the posterior mean of attributes estimates and the number of households having a statistically significant estimate tested by 95\% HPD regions, with the number of households for which HPD is positive and negative. Investigating these estimation results, we can notice that the number of households who have statistically significant response to the attributes differs from attribute to attribute, but stays big enough that the assumption of heterogeneity can be verified and valid.

\begin{table}[H] 
\caption{Hierarchical Model Estimation Results\label{tab4}}
\newcolumntype{C}{>{\centering\arraybackslash}X}
\begin{tabularx}{\textwidth}{CCCCCC}
\toprule
\textbf{}	& \textbf{Posterior Mean}	& \textbf{S.D.} & \textbf{HPD} & \textbf{(+)} & \textbf{(-)}\\
\midrule
Constant & -13.87 & 70.900 & 93 & 36 & 57 \\
S.A.A.   & 0.311  & 1.636  & 93 & 58 & 35 \\
Bleach   & 0.759  & 3.303  & 94 & 59 & 35 \\
Package  & 0.632  & 4.244  & 95 & 58 & 37 \\
g/30l    & -0.009 & 0.474  & 94 & 51 & 43 \\
net-w    & 1.195  & 10.311 & 95 & 57 & 38 \\
\bottomrule
\end{tabularx}
\end{table}

Regarding the average of estimates for the attributes, the highest is 1.195 of net-w (net weight of the detergent in the box). The lowest is -0.009 for g/30l (gram of detergent for 30 liters of water). In the case of g/30l, if consumer has to use more detergent for 30 liters, he will tend to have less utility for the detergent Product with such a type of attribute.

\section{Discussion}

In our approach, we followed the framework proposed by Kamakura and Russell (K-R); however, in our model, there are a few key advancements that need to be highlighted.

Unlike K-R's logit specification, our model leans on a probit framework, where the random variable in the utility equation follows a normal distribution.

A significant distinction from the K-R model is that our approach accounts for heterogeneity at the individual consumer level rather than within consumer segments. This shift means moving from a finite mixture to a continuous one, with ``s'' envisaged to approach ``h'' in magnitude.

The concept of product value is especially relevant, where it serves as a diagnostic instrument to assess the overall performance of a product, incorporating both its tangible and intangible aspects. An uncomplicated ranking of products based on market shares does not necessarily illuminate the explanations behind a product's particular performance level.

In the grand scheme of the environment, laundry detergents may seem a trivial factor. Yet, they harbor the significant possibility to enhance the sustainability quotient of our daily habits. By transitioning to environmentally conscious substitutes, embracing sustainable laundry habits, and critically examining packaging options, we can collectively contribute to the safeguarding of our planet for the generations yet to come.


\end{document}